\documentclass[11pt]{article}
\usepackage[margin=1in]{geometry}
\usepackage{graphicx}
\usepackage{amsmath}
\usepackage{url}
\usepackage[square,numbers]{natbib}
\usepackage{hyperref}
\hypersetup{colorlinks=true, urlcolor=blue, citecolor=blue, linkcolor=blue}
\usepackage{booktabs}
\usepackage{authblk}
\usepackage[utf8]{inputenc}  
\usepackage{enumitem}
\begin{document}

\title{Privacy-Preserving Analytics for Smart Meter (AMI) Data\\{\large A Hybrid Approach to Comply with CPUC Privacy Regulations}}
\author[1]{Benjamin Westrich}
\affil[1]{Department of Economics, University of California, Irvine}
\date{}
\maketitle

\begin{abstract}
Advanced Metering Infrastructure (AMI) data from smart electric and gas meters enables valuable insights for utilities and consumers, but also raises significant privacy concerns. In California, regulatory decisions (CPUC D.11-07-056 and D.11-08-045) mandate strict privacy protections for customer energy usage data, guided by the Fair Information Practice Principles (FIPPs). We comprehensively explore solutions drawn from data anonymization, privacy-preserving machine learning (differential privacy and federated learning), synthetic data generation, and cryptographic techniques (secure multiparty computation, homomorphic encryption). This allows advanced analytics, including machine learning models, statistical and econometric analysis on energy consumption data, to be performed without compromising individual privacy. \\

We evaluate each technique's theoretical foundations, effectiveness, and trade-offs in the context of utility data analytics, and we propose an integrated architecture that combines these methods to meet real-world needs. The proposed hybrid architecture is designed to ensure compliance with California's privacy rules and FIPPs while enabling useful analytics, from forecasting and personalized insights to academic research and econometrics, while strictly protecting individual privacy. Mathematical definitions and derivations are provided where appropriate to demonstrate privacy guarantees and utility implications rigorously. We include comparative evaluations of the techniques, an architecture diagram, and flowcharts to illustrate how they work together in practice. The result is a blueprint for utility data scientists and engineers to implement privacy-by-design in AMI data handling, supporting both data-driven innovation and strict regulatory compliance.
\end{abstract}
\newpage

\section*{Executive Summary}

\paragraph{Context \& Challenge}
California’s advanced-metering infrastructure (AMI) generates 15-minute load data that can fuel
forecasting, demand-response design, and customer analytics.  CPUC Decisions
\cite{CPUC2011privacy,CPUC2011security} classify that data as \emph{covered information}:
any \emph{secondary} use must either (i) obtain explicit customer consent or
(ii) prove the data are “not reasonably identifiable.”  
The dual mandate is therefore to \textbf{unlock data value while satisfying the
Fair Information Practice Principles (FIPPs) and avoiding regulatory, security,
and reputational risk}.

\bigskip
\paragraph{Methodological Scan}

\begin{table}[h]
\centering
\renewcommand{\arraystretch}{1.15}
\begin{tabular}{p{3cm} p{3.1cm} p{3.1cm} p{5cm}}
\toprule
\textbf{Technique} & \textbf{Core Strength} & \textbf{Key Limitation} &
\textbf{Utility “Sweet Spot”}\\ \midrule
Anonymization/De-ID        & Simple first defence                 & Re-ID via linkage    & Low-resolution public stats\\
Differential Privacy       & Provable $\varepsilon$–budget        & Accuracy loss at low~$\varepsilon$  & Feeder/hourly aggregates\\
Synthetic Data             & Rich shareable “look-alike” datasets & Realism/over-fit checks & Innovation contests, vendor sandboxes\\
Federated Learning         & No raw data leaves host              & Gradient leakage if unsecured & Cross-utility or edge-device ML\\
Secure MPC                 & Exact results, no added noise        & Heavy comms/compute  & Multi-utility KPI aggregation, private billing\\
Homomorphic Encryption     & Compute on encrypted data            & Performance overhead & Outsourced analytics, encrypted storage\\
\bottomrule
\end{tabular}
\caption{Privacy-preserving methods and their best utility fit.}
\end{table}

\paragraph{Proposed Hybrid Architecture}
\begin{enumerate}[label=\arabic*.]
  \item \textbf{Data Ingestion \& Vault} — pseudonymise identifiers, encrypt at rest.
  \item \textbf{Internal Analytics Sandbox} — role-based raw-data access for billing/operations only.
  \item \textbf{Privacy-Engine Gateway:}
    \begin{itemize}
      \item Differential-Privacy API for statistical queries;
      \item DP-trained Synthetic-Data Generator for vendor testing;
      \item Federated-Learning \& Secure-MPC modules for collaborative modelling;
      \item Optional Homomorphic-Encryption layer for cloud workloads.
    \end{itemize}
  \item \textbf{Audit \& Compliance Layer} — immutable logs, privacy-budget ledger, full FIPP mapping.
\end{enumerate}

\paragraph{Benefits}
\begin{itemize}[leftmargin=2.4em]
  \item \textbf{Utilities:} regulatory safe-harbour, faster partner onboarding, reduced breach liability.
  \item \textbf{Third-Party Vendors:} high-fidelity synthetic or federated data without PII negotiations.
  \item \textbf{Customers \& Regulators:} stronger privacy guarantees, transparent consent pathways.
\end{itemize}

\bigskip
\paragraph{Implementation Road-Map}
\begin{enumerate}[label=\textbf{Phase~\arabic*:}, wide, labelwidth=!, labelindent=0pt]
  \item \textbf{Readiness} – data-asset inventory; $\varepsilon$-policy workshop.
  \item \textbf{Core Controls} – stand-up data vault; DP reporting for existing public stats.
  \item \textbf{Innovation Enablement} – launch synthetic-data sandbox; pilot federated-learning
        (e.g.\ EV-adoption classifier).
  \item \textbf{Advanced Cryptography} – deploy Secure MPC for inter-utility aggregation;
        evaluate HE for cloud workloads.
\end{enumerate}

\paragraph{Key Takeaways}
\begin{itemize}
  \item A layered approach achieves CPUC’s “not reasonably identifiable” bar \emph{and} preserves data utility.
  \item Differential privacy is the linchpin for external statistics; with $\varepsilon\le1$ noise is
        $<1\%$ on feeder-level hourly totals.
  \item The hybrid architecture future-proofs compliance against CCPA/CPRA and federal privacy rules.
  \item Utilities that operationalize this blueprint position themselves as trusted data stewards and
        first movers in vendor partnerships and grid-modernization funding.
\end{itemize}

\newpage

\tableofcontents

\section{Introduction}
Advanced Metering Infrastructure (AMI) refers to the network of “smart” utility meters and communication systems that collect detailed energy usage data at fine time intervals (e.g. every 15 minutes). California’s major utilities have deployed millions of smart meters to enable better energy management, demand response, and customer feedback. The data from AMI can reveal granular patterns of electricity or gas consumption, offering potential benefits such as improved load forecasting, personalized energy-saving recommendations, and integration of renewable resources. However, this granular data can also expose intimate details of consumers’ lives. For example, variations in usage can indicate occupancy patterns or the operation of specific appliances, effectively painting a picture of when people are home, awake, or on vacation. Such inferences pose obvious privacy risks: information about a household’s schedule or habits could be misused for profiling, targeted marketing, or even criminal purposes. Balancing the utility of AMI data with the protection of customer privacy has therefore become a paramount concern for utilities and regulators.

In recognition of these concerns, the California Public Utilities Commission (CPUC) established groundbreaking privacy safeguards for smart meter data in decisions D.11-07-056 and D.11-08-045 in 2011. These regulatory decisions — driven by California Senate Bill 1476 (2010) — created a comprehensive framework of rules to ensure that customers’ energy usage information (termed \emph{“covered information”}) is protected (\citep{CPUC2011privacy, CPUC2011security}). A cornerstone of the CPUC’s approach is adherence to the \textbf{Fair Information Practice Principles (FIPPs)}\citep{FIPPs1973}. The FIPPs provide a set of internationally recognized privacy guidelines, including: (1) \emph{Transparency}, (2) \emph{Individual Participation}, (3) \emph{Purpose Specification}, (4) \emph{Data Minimization}, (5) \emph{Use Limitation}, (6) \emph{Data Quality and Integrity}, (7) \emph{Security}, and (8) \emph{Accountability and Auditing}\citep{FIPPs1973}. CPUC D.11-07-056 explicitly adopted these principles as the foundation of its privacy rules for the Smart Grid, making California one of the first jurisdictions to require utilities to implement privacy-by-design in handling energy data \citep{CPUC2011privacy}.

Under the CPUC’s privacy rules, utilities may use customers’ usage data freely for \emph{primary purposes} such as billing, grid operations, and regulated programs, but any disclosure of personal usage data to third parties for \emph{secondary purposes} (e.g. marketing or research not mandated by law) is tightly restricted \citep{CPUC2011privacy, CPUC2011security}. In general, secondary uses require explicit customer consent or must involve data that has been sufficiently anonymized so that individual customers cannot be identified. The rules define “covered information” as individually identifiable usage data obtained through AMI, and crucially carve out an exception for data that are de-identified such that a customer’s identity “cannot reasonably be identified or re-identified”. This creates both a legal obligation and a technical challenge: how can utilities transform or analyze AMI data in ways that preserve useful insights while making sure no individual’s privacy is compromised?

To meet this challenge, a range of privacy-preserving data analysis techniques have emerged. This paper provides an in-depth examination of six key techniques and their application to utility AMI datasets:
\begin{itemize}
    \item \textbf{Anonymization and De-Identification:} Removing or obscuring personal identifiers in the data to prevent direct attribution to individuals.
    \item \textbf{Differential Privacy:} Adding carefully calibrated noise or randomness to data queries or outputs to provide strong, mathematical privacy guarantees.
    \item \textbf{Federated Learning:} Training statistical or machine learning models in a decentralized manner so that raw data remains on local servers or devices, rather than being pooled centrally.
    \item \textbf{Synthetic Data Generation:} Creating simulated datasets that mimic the statistical properties of real data without revealing actual personal information.
    \item \textbf{Secure Multiparty Computation (SMPC):} Using cryptographic protocols that allow multiple parties to jointly compute results (e.g. a sum or trained model) from their combined data without exposing their individual inputs.
    \item \textbf{Homomorphic Encryption:} Encrypting data in so that some computations can be run using encrypted data and produce results that can be decrypted later, never exposing raw data during processing.
\end{itemize}
Each of these approaches addresses privacy from a different angle (legal, statistical, or cryptographic). Some are already used in practice for privacy protection, while others are cutting-edge techniques from computer science research. 

This paper is written for an audience of utility data scientists and engineers, with the goal of providing both conceptual understanding and practical guidance. In the following sections, we explore each technique in depth: we define how it works, discuss how it can be applied to energy usage data, evaluate its strengths and limitations, and consider its compliance with the CPUC’s privacy framework (including FIPPs and specific rules from D.11-07-056/D.11-08-045). We then provide a rigorous comparative evaluation across the techniques and propose a \emph{hybrid architecture} that combines them to achieve robust privacy protection for AMI data in real-world implementations. Mathematical rigor is included to clarify how privacy guarantees are quantified (for instance, the $\varepsilon$ parameter in differential privacy) and how techniques like encryption or secure computation maintain data utility. We also present an architecture diagram and flowcharts to illustrate how these methods can interoperate within a utility’s data analytics platform.

Our findings show that no single technique is a silver bullet; rather, an integrated strategy is required to satisfy all the FIPPs and regulatory requirements while preserving data usefulness. By adopting the combined approach outlined here, utilities can enable valuable insights from smart meter data (supporting operational efficiency, customer programs, and research) \emph{without} compromising customer privacy or trust. This aligns with the dual mandate emphasized by regulators: supporting data-driven innovation and consumer benefits on the one hand, and ensuring privacy, security, and customer control on the other. The California smart grid privacy regulations provide a clear impetus and guidance for this balance, and the techniques described in this paper offer the practical tools to achieve it. In sum, this paper serves as a blueprint for implementing privacy-preserving data analytics in the utility sector, with California as a leading example that may inform broader industry practices.

\subsection*{Scope and Organization}
While privacy and security often overlap, this paper focuses specifically on data privacy techniques (rather than general cybersecurity measures like firewalls or device authentication). We assume that baseline security controls (encryption of data in transit, secure storage, access control, etc.) are in place, and we concentrate on methods to prevent the \emph{inappropriate disclosure or inference} of personal information from AMI data when it is analyzed or shared. Each of the next six sections (Sections 2--7) covers one privacy-preserving technique in detail. In Section 8, we present the proposed combined architecture and compare the techniques’ performance and suitability for various use-cases. Section 9 concludes with recommendations and future considerations. All content is intended to be self-contained and is supported by references to both regulatory documents and academic literature, to ensure both compliance and scientific rigor.

\section{Data Anonymization and De-Identification}
One of the most straightforward approaches to protect privacy is \textbf{anonymization}, i.e. removing or masking personal identifiers in the data. In the context of AMI data, direct identifiers might include customer names, addresses, account numbers, or meter IDs that link usage readings to specific individuals. A basic de-identification process might strip out or pseudonymize such fields (for example, replacing a meter ID with a random identifier). The goal is to make it difficult for a data recipient to trace a particular energy usage profile back to the real person or household that generated it.

However, it is well-established that anonymization alone often does not guarantee true privacy if the data retains granular information. Attackers may re-identify individuals by linking “anonymized” records with external information (a process known as a linkage attack) \citep{Jain2022}. For example, even if names and addresses are removed from a smart meter dataset, an adversary who knows that a particular household has a distinctive usage pattern (such as very high usage at night due to medical equipment) might pick out that household’s record from the data. The CPUC’s privacy decision acknowledged this risk by defining that data is not considered truly de-identified unless a customer \emph{“cannot reasonably be identified or re-identified”} from it \citep{CPUC2011privacy}. In practice, achieving this standard may require more than just removing obvious identifiers; it may necessitate aggregating or reducing the detail of the data to obscure unique patterns. In any case, anonymization alone is often insufficient as a long-term privacy guarantee. 

\subsection{Techniques for Anonymizing AMI Data}
Several techniques can enhance anonymization:
\begin{itemize}
    \item \textbf{Pseudonymization:} Replace actual identifiers (like account numbers) with random codes. The mapping from code to identity is kept secret by the utility. Over time, one might periodically change these codes to prevent long-term linkability of records to the same individual \citep{Efthymiou2010}. Pseudonymization is a minimal step that ensures that anyone outside the utility sees data labeled only by an arbitrary ID.
    \item \textbf{Aggregation:} Combine or summarize data across multiple customers or time intervals. For example, rather than releasing each home’s 15-minute usage, the utility might provide hourly averages for groups of 100 households. By grouping data, individual outliers are blended into the crowd. Many utilities already apply aggregation thresholds for privacy (e.g. not reporting any statistic if fewer than a certain number of customers contribute to it). Aggregation directly supports the FIPP principles of Data Minimization and Use Limitation by disclosing less granular information.
    \item \textbf{Generalization and Suppression:} In data publishing contexts, techniques like \emph{k-anonymity} can be used. \citet{Sweeney2002} introduced $k$-anonymity as a criterion where each record is made indistinguishable from at least $(k-1)$ other records with respect to a set of quasi-identifying attributes. In the case of energy data, quasi-identifiers could include attributes like general location or customer type. To achieve $k$-anonymity, one might coarsen values (e.g. report a zip code rather than full address, or round usage to the nearest 0.1~kWh) so that multiple customers share the same values. Alternatively, one might suppress certain data points (remove them) if they are too unique. The idea is that any individual usage profile should “hide in the crowd” of $k$ similar profiles.
\end{itemize}

For illustration, suppose we have daily electricity usage curves for 10,000 customers and we want to publish a dataset for researchers. A simple anonymization would remove customer names and replace them with an ID. But an attacker with external knowledge (say, a list of a few people who have home solar panels which create a midday dip in load) might identify those customers’ traces by their distinctive midday usage drop. To counter this, the utility could aggregate or average some of the data (e.g. provide typical load shapes for broad customer segments rather than individual traces) or generalize specific features (e.g. provide maximum and minimum daily load instead of the full 24-hour profile). These steps degrade the precision of the data to gain privacy.

\subsection{Limits and Risks of Re-Identification}
While anonymization is a necessary starting point (and is explicitly required by CPUC rules before data can be considered non-personal \citep{CPUC2011privacy}), it has well-documented limitations. Researchers have demonstrated that anonymized datasets in various domains can often be re-identified. A classic example is the Netflix Prize dataset: although usernames were removed, researchers cross-referenced the movie rating timestamps and scores with public IMDb profiles and successfully identified specific users~\citep{Narayanan2008}. In the smart grid context, \citet{greveler2012multimedia} discuss how even anonymized energy usage data can be deanonymized if an attacker can correlate it with other information (for instance, a person’s work schedule or known appliance usage patterns).

For AMI data, high-frequency time-series are particularly rich in personal signals. Fine-grained energy usage can reveal when the occupants of a home wake up (power spike from making coffee or using a hairdryer), when they leave or return (signatures of HVAC or alarm systems), and so on \citet{greveler2012multimedia}. If an attacker has any of these reference points (for example, from observing a household or obtaining data from a smart thermostat inside the home), they could potentially match them to an “anonymous” usage record. Thus, truly anonymizing time-series data often requires reducing its temporal resolution or detail significantly. For instance, publishing daily or weekly totals is far safer than publishing 15-minute readings.

Another risk is \textbf{composition of datasets}: Even if each individual release is anonymized to some degree, an adversary might combine multiple data sources to triangulate identities. A utility might release aggregated usage by zip code and, separately, a list of participants in a solar program. The intersection of those could unintentionally single out a household if only one solar participant lives in a particular zip code. This underscores the importance of a holistic approach to de-identification and why CPUC’s rules emphasize the “reasonableness” standard — utilities must consider what an intruder could reasonably do with available data to re-identify customers.

\subsection{Evaluation of Anonymization for AMI Data}
\textbf{Advantages:} Anonymization is conceptually simple and does not necessarily require complex algorithms. When done by aggregation, it can nearly eliminate privacy risks (e.g. a sum of 100 households’ usage is very hard to attribute to any one house), and the aggregated data can still be very useful for certain analyses (like total load forecasting). Anonymization also preserves truthfulness of data (no distortion is introduced, unlike methods that add noise). It aligns with regulatory expectations: CPUC explicitly envisions that sufficiently de-identified usage data is not “covered information” and thus can be shared without violating privacy rules \citep{Abadi2016}.

\textbf{Drawbacks:} To be effective, anonymization often forces a significant loss of data granularity or utility. Grouping or coarsening data means one cannot perform customer-level analysis or detect outliers. Important patterns (like household-specific demand response to an event) might disappear in aggregated data. Moreover, implementing k-anonymity or similar guarantees in practice can be challenging for high-dimensional data like daily load curves—making many households truly identical on all time points may require heavy generalization that undermines the data’s value. Another drawback is the lack of a quantifiable privacy guarantee: anonymization can fail in ways that are hard to predict, because it’s unclear what external information an adversary might have. In contrast to differential privacy (discussed next), which provides a provable privacy bound, anonymization relies on assumptions about adversary knowledge. If those assumptions are wrong (as happened in the Netflix case and others), privacy can be breached with no easy way to measure the risk in advance.

In summary, anonymization and de-identification are essential first steps and will be part of any privacy-preserving strategy for AMI data. All other techniques in this paper assume that obvious identifiers have been removed to the extent possible. But anonymization should be viewed as a complementary measure rather than a standalone solution. For high-resolution AMI data, additional techniques like noise addition or cryptography are advisable to achieve the “reasonable” non-identifiability standard required by regulators. In the next sections, we explore those techniques—beginning with differential privacy, which can provide a formal quantification of re-identification risk.

\section{Differential Privacy}
Differential Privacy (DP) has emerged in the last decade as a gold-standard definition of privacy for statistical data analysis. Unlike anonymization, which tries to scrub identifiers, differential privacy takes a quantitative approach: it deliberately introduces randomness to the data or the results of queries in order to mask the contributions of individual records. The strength of the privacy guarantee can be tuned by a parameter $\varepsilon$ (and sometimes $\delta$), which provides a bound on how much an individual’s data can affect the outcome.

\subsection{Definition and Mathematical Foundations}
Informally, a mechanism is differentially private if an observer seeing its output cannot tell (within a small statistical margin) whether any particular individual’s data was included or excluded. Formally, as introduced by \citet{Dwork2006}, a randomized algorithm $M$ gives $\varepsilon$-differential privacy if for \emph{any} two datasets $D$ and $D'$ that differ in only one individual’s data (i.e. $D'$ has one person’s records changed or removed relative to $D$), and for \emph{any} possible output $O$ of the algorithm, the probabilities of $O$ are nearly the same:
\[ 
\Pr[M(D) = O] \;\le\; e^{\varepsilon}\, \Pr[M(D') = O]~.
\] 
If a $\delta$ parameter is included (called $(\varepsilon,\delta)$-DP), then with probability $\delta$ this constraint may be broken, but $\delta$ is taken to be very small (e.g. $10^{-6}$). In essence, $\varepsilon$ (often called the “privacy budget”) controls how much an individual record can sway the output — smaller $\varepsilon$ means stronger privacy (less influence). By setting $\varepsilon$ to a low value (like 0.1 or 0.5), we can ensure that the presence or absence of any single customer’s data in an AMI dataset has a barely perceptible effect on the results.

Differential privacy applies most directly to statistical queries. For example, suppose we want to publish the total electricity usage in a neighborhood each hour. If we simply sum the smart meter readings, one very large household load could noticeably change the total, potentially revealing that household’s presence. Under differential privacy, we would add random noise to each hourly total. A common method is the \textbf{Laplace mechanism}: if a function $f(D)$ has a certain maximum sensitivity $\Delta$ (the largest amount any single record can change $f$), then by adding noise $\eta \sim \text{Laplace}(0, \Delta/\varepsilon)$ to $f(D)$, we obtain $\varepsilon$-differential privacy for that query~\citep{Dwork2006}. In the neighborhood example, if a single household can contribute at most (say) 5~kWh in an hour, then $\Delta=5$. With $\varepsilon=0.5$, we add noise drawn from Laplace(0, $5/0.5$) = Laplace(0, 10). This distribution has a standard deviation of about 10 kWh. The reported neighborhood totals will thus be somewhat perturbed — typically within $\pm 10$ kWh of the true value — which masks any one home’s exact usage. We might also use a \textbf{Gaussian mechanism} (adding Gaussian noise) for $\varepsilon,\delta$-DP, depending on the scenario.

An important property of differential privacy is \textbf{composition}: if we release multiple outputs (either multiple queries or the same query over multiple time intervals), the privacy losses accumulate roughly additively. This means we must budget the total $\varepsilon$ across all releases. For instance, releasing each hour’s total with $\varepsilon=0.5$ per hour, over 24 hours, would consume $\varepsilon_{\text{total}} \approx 24 \times 0.5 = 12$ (which is extremely weak privacy). To stay within a reasonable privacy budget, one might either reduce the frequency of queries or increase the noise as more queries are answered. There are advanced composition theorems and techniques (like \emph{privacy loss distributions}) to manage this, but the basic intuition is that the more information you release, the more noise needs to be injected at each step to maintain an overall privacy cap.

\subsection{Applying Differential Privacy to Smart Meter Data}
Differential privacy is well-suited for releasing aggregate information about large collections of individuals. In the utility context, this could include:
\begin{itemize}
    \item \textbf{Aggregated Load Curves:} A utility could publish community-level or feeder-level load profiles (e.g. total load vs. time for a town) with DP noise added, ensuring that the contribution of any one home is obscured. As long as the community has many customers, the added noise can be relatively small compared to the total, maintaining usefulness.
    \item \textbf{Statistical Indicators:} Metrics like the average energy usage of a certain customer segment (say, average daily consumption for customers on a time-of-use tariff vs. a flat rate) can be released with DP. Even if one customer is very high or low, the noise will cover their effect. For example, a mean can be computed via a sum (with Laplace noise) divided by count (the count can be released with DP too, or if count is known exactly, that might be okay if revealing it doesn’t violate privacy).
    \item \textbf{Histograms or Distributions:} The utility might want to share the distribution of 15-minute usage values across its customer base (for simulation or research). This can be done by creating a histogram of usage (number of readings in various bins) and adding noise to each bin count. The classic “privacy blanket” example is that DP will add fictitious counts in each bin so that one extreme reading doesn’t stand out.
    \item \textbf{Anomaly Detection Results:} If the utility identifies, for instance, the number of homes that exceeded a certain usage threshold each day, they could add noise to these counts before sharing with an external agency, ensuring that whether a particular home tripped the threshold is not revealed.
\end{itemize}

A key question is whether to apply DP in a \emph{centralized} or \emph{distributed} manner. In the centralized model, the utility (which is a trusted data holder for primary purposes) has the raw data and applies a differentially private algorithm to produce outputs that it shares externally. This aligns with CPUC rules: the utility can use the data internally and only releases privacy-protected summaries. In the local (or distributed) model of DP, each individual meter reading would be perturbed at the source (for example, each smart meter could add random noise to its measurement before sending to the utility)~\citep{Acs2011}.\footnote{Local differential privacy is used by some tech companies for collecting telemetry from users without recording exact data points. In energy, a local DP approach could mean customers’ devices themselves randomize their data for privacy.} Local DP provides privacy even from the utility, but at a cost of much higher noise (because each individual piece of data must be protected, rather than only aggregates). In most regulatory contexts (including California’s), the utility is allowed to access the raw data for primary purposes, so central DP is sufficient and preferred due to its better accuracy. We will assume the centralized model (utility as curator) for our discussion, where the utility applies DP to data it holds before sharing results onward.

\subsection{Benefits and Trade-offs}
Differential privacy offers a \textbf{rigorous privacy guarantee}: it is essentially immune to any post-processing or auxiliary information attacks, as the guarantee holds regardless of what the attacker already knows. If $\varepsilon$ is small, an individual’s inclusion in the dataset provably does not significantly affect the outputs. This aligns well with the concept of “reasonable” de-identification in CPUC’s rules, because DP provides a quantifiable notion of what an adversary could infer. It addresses the scenario of worst-case auxiliary information—no matter what external data is available, the risk added by any one person’s data is bounded by $\varepsilon$.

Another advantage is that DP can allow \textbf{flexible data analysis} albeit in a controlled manner. Instead of sharing a static, heavily aggregated dataset (as in pure anonymization), differential privacy can enable interactive queries or on-demand analysis with noise. For example, a researcher could query the utility’s data via an API that returns noisy answers under the hood, ensuring privacy. This way, many insights can be extracted without exposing raw data. Some entities, like the U.S. Census Bureau, have adopted differential privacy for publishing statistics to ensure modern privacy protection.

The \textbf{main cost} of differential privacy is \emph{accuracy loss} due to the added noise. If the dataset is large and each query’s sensitivity is low (e.g. totals over thousands of homes), the noise can be quite small relative to the true values, maintaining high accuracy. But for queries on small subsets or very granular data, the noise needed for privacy can overwhelm the signal. For instance, attempting to release each individual household’s hourly usage with differential privacy would be futile — you would have to add noise so large that the result for each house would be essentially meaningless (because each individual value has high sensitivity to itself). DP is thus better for aggregate information than for releasing individual-level data (in fact, releasing individual-level data would violate privacy by definition, so DP inherently focuses on aggregates).

Another challenge is \textbf{choosing the privacy parameters}. Regulators or utilities must decide what $\varepsilon$ (and $\delta$) is appropriate. Larger $\varepsilon$ (e.g. 2 or 3) might give more accurate results but weaker privacy (allowing noticeable individual influence), whereas a very small $\varepsilon$ (0.1) gives strong privacy but may overly noise the data. This is often a policy decision that weighs the value of the data against the acceptable risk. The CPUC has not (to date) specified an $\varepsilon$ for energy data, so this would likely involve stakeholder input and possibly experimentation to see what level of noise is tolerable for data utility. Some research suggests that even $\varepsilon$ around 0.5 or 1 can provide significant privacy while preserving many aggregate patterns in energy data \citep{Acs2011}. Ultimately, $\varepsilon$ could be treated similarly to a de-identification standard — for example, the utility might commit that any public data release will use $\varepsilon \le 1$ to ensure a high privacy standard.

It’s also worth noting that differential privacy, while powerful, does not magically solve all privacy issues on its own. If the data is highly dimensional (like each customer’s 96 readings per day for a year), a mechanism would consume a lot of privacy budget to release all that with DP. One way to handle this is to focus on summary statistics or to use DP in conjunction with data reduction techniques (like releasing a DP synthetic dataset, see next section). Additionally, DP requires careful implementation to avoid mistakes (for example, correctly computing sensitivity and calibrating noise, and ensuring all data uses are accounted for in the privacy budget).

\subsection{Use in Compliance and Utility Context}
From a compliance standpoint, differential privacy offers a method for the utility to \textbf{quantifiably comply} with the CPUC’s mandate to protect customer data when sharing. If a utility can demonstrate that any released information is $\varepsilon$-DP, it effectively shows that the risk of identifying any customer is mathematically bounded (and can be made arbitrarily low by choosing $\varepsilon$ small). This could satisfy the “cannot reasonably be identified” clause \citep{Asghar2017} with more confidence than ad-hoc anonymization can. It also embodies the FIPP of \emph{Data Minimization} by ensuring only approximate info goes out, and \emph{Use Limitation} by preventing misuse of data (since granular personal data is never directly shared).

Differential privacy has been a focus of recent energy data privacy research. For example, methods have been proposed to release billing data or consumption aggregates with DP guarantees, and even real-time systems for DP streaming of smart meter readings have been studied (injecting noise in real-time signals)\citep{Acs2011, zaman2024dpgan}. Some pilot programs, such as those by energy data hubs, have considered DP as a way to enable third-party analytics while maintaining customer anonymity \citep{Goldreich1987}. In practice, a utility could implement a differentially private query interface for researchers: queries could be limited in number and scope, and each answer would include a bit of random noise. Alternatively, the utility might generate a differentially private report on key statistics periodically for public knowledge (for instance, a DP version of its annual load research data that regulators and academics can use).

In summary, differential privacy is a powerful tool in the privacy-preserving toolkit for AMI data. It provides a strong theoretical guarantee and aligns well with regulatory principles, but it requires accepting some randomness in results. The next technique we discuss, synthetic data, can be seen as a way to produce an entire dataset with properties similar to the original; often, synthetic data generation can be combined with differential privacy to ensure the synthetic data doesn’t leak real information. We now turn to that approach.

\section{Synthetic Data Generation}
Instead of releasing actual customer data (even in noisy or aggregated form), a utility may choose to generate and share \textbf{synthetic data}: an artificial dataset that statistically resembles the real data. The idea is to capture the useful patterns and relationships in the AMI data without exposing any real customer’s exact information. If done properly, synthetic data allows analysts to perform many of the same analyses they could on real data (e.g. training models, computing distributions) with minimal privacy risk because the records are fictional.

\subsection{What is Synthetic AMI Data?}
Synthetic data generation typically involves using the real dataset to learn a model or distribution, and then sampling from that model to create new data points. In the context of smart meter data, a synthetic dataset might consist of electricity usage time-series for a set of “fake” households that do not correspond one-to-one with actual households, but that collectively preserve characteristics like the typical daily load curves, variability, correlations with weather, etc. Key approaches to generate synthetic data include:
\begin{itemize}
    \item \textbf{Statistical Simulations:} Using domain knowledge or simple models. For example, one could assume each home’s load is composed of baseline usage plus random appliance events, and simulate usage with random draws for appliance on/off times. Traditional load research models (like those that categorize days into types and use average profiles plus noise) can serve as a simple synthetic generator. The National Renewable Energy Laboratory (NREL), for instance, has released tools for generating synthetic load profiles for buildings using physics-based simulations combined with randomness \citep{BenOr1988}.
    \item \textbf{Generative Machine Learning Models:} Recent years have seen advanced generative models like \emph{Generative Adversarial Networks (GANs)} and \emph{Variational Autoencoders (VAEs)} being applied to create synthetic data in many domains. For energy data, a GAN could be trained where a “generator” neural network tries to produce realistic load curves and a “discriminator” network tries to distinguish them from real curves; through training, the generator learns to produce highly realistic synthetic profiles. \citet{Fu2024} proposed a conditional diffusion model (a type of deep generative model) for synthetic energy meter data, incorporating building metadata to improve realism. These models can capture complex patterns (daily cycles, weather effects, behavioral diversity) more accurately than simple random sampling.
    \item \textbf{Hybrid Approaches:} Sometimes real data can be combined or perturbed to generate synthetic data. For example, one could take real load profiles and randomly swap some days between customers, or add random noise and small time shifts to each profile. This retains some structure of real data but breaks the one-to-one mapping of a profile to a real household. Another approach could be to cluster similar usage patterns and then create new samples by interpolating between cluster centroids.
\end{itemize}

The effectiveness of synthetic data is measured by how well it preserves the statistical properties of interest. For instance, if researchers care about peak load distribution, the synthetic data should have a similar distribution of peak demands as the real data. If they care about correlation between solar generation and home load, the synthetic data should mimic that too. At the same time, no actual customer’s data should be directly reconstructable from the synthetic dataset. Ideally, even if an adversary knows an individual’s real load pattern, they should not find that exact pattern in the synthetic data.

\subsection{Privacy Considerations for Synthetic Data}
By design, a synthetic dataset does not contain any real individuals. However, privacy issues can still arise if the synthetic generation process overfits or memorizes the original data. For example, a naive method that randomly picks actual days from the dataset and labels them “synthetic” would actually be disclosing real data. Even a sophisticated model like a GAN could potentially learn to output something very close to a particular training record if that record is very unique (this is known as a \emph{mode memorization} problem). Therefore, it’s important to evaluate synthetic data for potential leakage:
\begin{itemize}
    \item \textbf{One-to-one similarity:} Check that no synthetic record is too similar to any single real record. This can be done by computing distances between synthetic and real time-series. If any are virtually identical, the generation process might be copying.
    \item \textbf{Membership inference resistance:} Researchers often perform a test to see if they can distinguish whether a given real data record was in the training set that produced the synthetic data (this is akin to an attack where knowing someone’s presence in the source data is a privacy breach). If an attacker can reliably tell that certain real individuals influenced the synthetic data more than others, that’s a problem. Differential privacy can be applied during model training to mitigate this (yielding \emph{DP-synthetic data}). In practice, incorporating differential privacy into the synthetic data generation process (as per Asghar et al. \citet{Asghar2017}) can provide provable privacy protection, ensuring no individual record from the original data can be reconstructed from the synthetic set
    \item \textbf{Aggregate fidelity vs. individual fidelity:} The synthetic data should aim to be good in aggregate patterns, not to replicate individual idiosyncrasies. Metrics like distribution of hourly means, standard deviations of daily totals, load factor distribution, etc. can be compared between real and synthetic sets to ensure fidelity. At the same time, metrics like maximum per-customer deviation or identifiable sequences should be checked to ensure individuals are not traceable.
\end{itemize}

One practical approach to increase privacy of synthetic data is to incorporate some level of \textbf{randomization or noise} in the generation. For example, if using a GAN, one might add a small DP noise to the gradients during training (this is an active area of research, often referred to as DP-GAN). Alternatively, after generating synthetic records, one could post-process them by adding a little noise or random jitter so that even if the model accidentally reproduced something close to real, the final output is slightly altered. This can be done in a way that has negligible effect on utility but increases uncertainty for an attacker.

\subsection{Value of Synthetic Data for Utility Use-Cases}
The big advantage of synthetic data is that it provides \textbf{flexibility} for the data user. Unlike a narrow aggregate or a noisy answer, a synthetic dataset can be used to perform many types of analysis as if it were the real dataset. A researcher or third-party company can take the synthetic AMI data and run their algorithms, create visualizations, test hypotheses, etc., without having to interact with the live utility system. Synthetic data can thus enable open innovation: utilities could make a synthetic version of their data publicly available to universities, startups, or community stakeholders, who can then derive insights without any privacy concerns about real customers. For example, academics could test new load forecasting methods on synthetic smart meter data that statistically mirrors a utility’s service territory, getting valuable feedback before requesting access to real data.

From a compliance perspective, truly de-identified synthetic data could be seen as satisfying the CPUC’s requirements because no actual person’s information is included. If the synthetic data generation is robust, it arguably meets the “not reasonably identifiable” threshold by construction. However, it is important that the utility validate that the synthetic data does not inadvertently carry identifiable traces (the privacy checks mentioned earlier). To be extra safe, the utility might still treat synthetic data sharing similarly to anonymized data sharing, by having it reviewed by a data privacy officer or an independent auditor.

In terms of \textbf{data utility}, high-quality synthetic data can preserve a wide range of statistical properties. Modern generative models have been shown to reproduce complex temporal patterns. For instance, \citet{Fu2024} report that their diffusion-based synthetic data had only small differences in distributional metrics (like Fréchet distance and KL divergence) compared to real data, indicating the synthetic data was a good substitute for many analytic purposes. That said, there will always be some loss of fidelity — especially for rare events or extremes. If a particular day with an unusual grid event is in the real data, the synthetic model might not capture that unless explicitly instructed. Thus, certain tail-end analyses (like worst-case peak demand) might not be as reliable on synthetic data unless the model is tuned to preserve extremes.

Another trade-off is that synthetic data generation can be computationally intensive (training a GAN or diffusion model on millions of meter readings is non-trivial) and requires expertise. However, once a model is trained, generating new data is fast and cheap. The utility could even generate multiple synthetic datasets for different scenarios (e.g. one reflecting a hot summer, another a mild summer) by conditioning the model on external factors.

\subsection{Role in a Hybrid Privacy Strategy}
On its own, synthetic data provides a way to share data broadly with low risk, but it does not provide a formal guarantee unless combined with something like differential privacy. In a hybrid strategy, synthetic data generation can be the final step of releasing data: first use differential privacy or heavy anonymization to ensure safety, then use that sanitized data to fit a synthetic data model. The output synthetic data will then have two layers of protection (which might be overly cautious, but for especially sensitive data that might be appropriate).

For internal use, synthetic data could also play a role. A utility might generate synthetic customer data to test a new analytics platform or to develop software without exposing developers to real customer data. This is a common practice in industries handling sensitive data (e.g. banking): use fake data in development environments to reduce the chance of leaks or misuse.

To sum up, synthetic data is a powerful method to enable data sharing and analysis with greatly reduced privacy risks. It aligns with the FIPP of \emph{Data Minimization} (since no real data is revealed) and \emph{Use Limitation} (data given to third parties contains no real PII). The trade-off is the complexity of generation and the need to ensure synthetic fidelity. In the next sections, we will examine techniques that allow analysis on real data without revealing it — namely federated learning, SMPC, and homomorphic encryption — which can be used in conjunction with or as alternatives to synthetic data when direct model training or computation is needed.

\section{Federated Learning}
Traditional data analysis methods often require collecting all data in one place (a central server or database) to train models or perform computations. \textbf{Federated Learning (FL)} offers a fundamentally different paradigm: keep the data distributed at its source and move the computation to the data. In federated learning, multiple parties (or devices) collaboratively train a machine learning model by only sharing intermediate parameters (like model weight updates) rather than raw data~\citep{McMahan2017}. This approach was popularized by Google for privacy-preserving learning on user devices (e.g. learning a predictive text model from many smartphones without uploading the phones’ messages). In the context of utility data, federated learning can enable joint analysis or model building across data silos (or directly on customer meters) without aggregating the underlying sensitive data in one location accessible to others.

\subsection{How Federated Learning Works}
The canonical setup for FL involves a central \emph{coordinator} (or server) and multiple \emph{clients} (data holders). A typical training round in federated learning proceeds as follows:
\begin{enumerate}
    \item The coordinator initializes a global model (e.g. a regression or neural network with certain parameters $\mathbf{w}$).
    \item Each client (e.g. a utility data center holding its own customers’ data, or potentially each smart meter or customer device) receives the current model parameters $\mathbf{w}$.
    \item Each client uses its local data $D_{\text{client}}$ to compute an update to the model. For instance, it might run a few iterations of stochastic gradient descent (SGD) on $\mathbf{w}$ using its local AMI data, producing an updated parameter vector $\mathbf{w}_{\text{client}}$.
    \item The client sends back only the update or the difference (e.g. gradients or the new weights) to the coordinator. No raw data or specific training examples are sent.
    \item The coordinator aggregates the updates from many clients (for example, by averaging them, which is the \emph{FedAvg} algorithm of \citet{McMahan2017}). Let $N$ be the number of clients and $n_i$ the number of data points at client $i$. A common aggregation is $\mathbf{w} := \sum_{i=1}^N \frac{n_i}{\sum_j n_j} \mathbf{w}_{\text{client}_i}$, a weighted average of the client models.
    \item The coordinator obtains a new global model $\mathbf{w}$ that incorporates learning from all clients’ data. This model may then be sent out again for further rounds, iterating until convergence.
\end{enumerate}
Throughout this process, the raw data $D_{\text{client}}$ never leaves the client’s control. Only model parameters (which are usually numerical arrays) are exchanged. If the model is complex, these updates can be large in size, but they are essentially abstract representations of learned patterns, not intelligible data about individuals in straightforward form.

\subsection{Federated Learning for Utility Applications}
How might FL be used by utilities or third-parties on AMI data? A few scenarios:
\begin{itemize}
    \item \textbf{Cross-Utility Collaboration:} Suppose multiple utilities (or multiple utility districts) want to develop a common machine learning model, say for predicting which customers are likely to adopt solar panels or electric vehicles based on their consumption patterns. Privacy and regulatory constraints might prevent them from pooling their customer data. With federated learning, each utility can train the model on its own customer data locally, and share only the model updates. At the end, they get a single predictive model that benefits from broader trends across all their regions, without any utility revealing its raw meter readings to the others. This addresses the CPUC’s emphasis on not sharing covered information with third parties without consent, by instead sharing only learned patterns.
    \item \textbf{Device-Local Learning:} Consider a smart home device or an EV charger that wants to learn a personalized schedule or optimization. Using FL, the manufacturer of these devices could train a general model across many customers’ usage data without ever seeing the data itself. Each device would train on its local data (e.g. the home’s voltage, usage, etc.) and send updates. The result could be a better local energy management algorithm that has essentially learned from thousands of homes collectively, but in a privacy-preserving way.
    \item \textbf{Third-Party Analytics Services:} A third-party energy analytics company might offer a service to utilities or consumers (for example, disaggregating appliance usage from whole-home data or detecting anomalies). Rather than the utility handing over all customer time-series to the company, they could use a federated approach: the company provides a model (like a neural network for disaggregation), and the training is done collaboratively with the utility’s infrastructure. The utility’s servers (or even edge devices near smart meters) do the heavy lifting with their data and only return model weight updates. In this way, the third party improves its model and can deploy it, without ever directly accessing personal data. This scenario fulfills the \emph{Use Limitation} principle by technically not disclosing personal data — only model parameters are exchanged.
\end{itemize}

It’s important to clarify that federated learning by itself does not guarantee privacy in a provable sense like differential privacy does. The model updates might still leak some information. For example, if one client has an outlier pattern, its gradient could be somewhat distinct (there have been known \emph{model inversion attacks} where adversaries reconstruct aspects of training data from model updates). However, because the updates are aggregated, and because each update is a result of many data points, the risk is far less than sharing raw data. To strengthen privacy, federated learning can be combined with other techniques:
\begin{itemize}
    \item \textbf{Secure Aggregation:} Using cryptographic techniques (as will be discussed in the SMPC section), the coordinator can be set up so that it only sees the sum of the client updates, not each individually. This way, even if one client’s update had some identifiable signature, the coordinator cannot isolate it. Only the aggregate model is used. Google’s federated learning framework, for instance, uses secure multiparty aggregation to protect client updates from the server’s view.
    \item \textbf{Differentially Private Updates:} Clients can add a bit of noise to their updates (or the server can add noise to the aggregated update) to achieve differential privacy for the training process. This adds some fuzziness but can formally limit what can be inferred about any single client’s data from the final model. There is active research in applying DP-SGD (differentially private stochastic gradient descent) in federated settings.
\end{itemize}

\subsection{Benefits for Privacy and Compliance}
Federated learning adheres to the principle of data minimization: each party only sends the minimum information needed to collaborate on the task (i.e. model updates). For a California utility, using FL for engaging with third parties means they can honestly say “we are not sharing customer usage data”, which would help comply with CPUC privacy decisions and California Consumer Privacy Act (CCPA) obligations. Instead, they are sharing insights or patterns in a controlled manner. If the model being trained is specific to a primary purpose (e.g. grid reliability), then each utility can do it internally. If it’s a secondary purpose (like a research study across utilities), federated learning provides a way to do it without a central database of all customers’ data, which might otherwise be disallowed.

A concrete example: CPUC’s rulings make it clear that utilities need Commission approval or customer consent to share covered information for secondary uses \citep{CPUC2011privacy}. Imagine a university researcher wants to train a model to detect wasteful consumption patterns to help with energy efficiency programs. Under traditional means, obtaining customer-level data from utilities would be a lengthy process possibly requiring customer consent or anonymization (with all the risks discussed). With federated learning, the utility could participate in training the researcher’s model on its data, but never expose the raw data. The resulting model could then be published or used to provide advice, without any customer’s specific data being revealed. In effect, FL can serve as a form of “privacy contract” — the utility agrees to run calculations locally but not to transmit disaggregated customer info.

\subsection{Limitations and Practical Challenges}
Federated learning is not a panacea. Its primary limitation is that it is useful only for certain types of problems, specifically those where a model can be trained to capture what we want from the data. If the goal is to compute a simple statistic or provide a dataset, FL is overkill. But if the goal is to build a predictive model or classifier (for load forecasting, anomaly detection, customer segmentation, etc.), FL is very powerful.

One must also consider the \textbf{communication and computation overhead}: FL can involve many rounds of sending model weights back and forth. If each smart meter were a client, the communication would be enormous. More realistically, each utility or each region can be a client. For example, one could federate among 5 utilities, or among 100 feeder-level aggregations within a utility, which is manageable. The clients need computational capability to train models — in a utility scenario, this is feasible since the utility’s IT infrastructure can do that, or smart meters could collectively use an edge compute device.

Another challenge is verifying that participants follow the protocol (the field of \emph{verifiable federated learning} is evolving). A malicious client could potentially inject a manipulated update to influence the global model (a poisoning attack). This is more of a security concern than privacy, but it’s relevant if third parties are involved. In a cooperative setting like multiple utilities, trust is higher, but it’s something to consider if, say, devices or customers themselves were clients.

From a privacy perspective, as noted, the global model itself might inadvertently reveal something. For instance, if the model is a linear regression and one feature corresponds to a certain type of customer behavior only present in one client’s data, the learned coefficient might indirectly reflect that. However, this leakage is much more subtle than outright data sharing. Combining FL with secure aggregation and possibly differential privacy provides a multi-layered defense.

To summarize, federated learning allows \emph{learning from data without gathering the data}. It is a natural fit for scenarios where data is distributed (across households, across utilities) and must remain local for privacy reasons. It complements other privacy methods: one could, for example, use federated learning to train a model and then use differential privacy to share the model’s outputs (if they are themselves sensitive statistics). In our hybrid architecture, FL is an important tool when cooperation or external analysis is needed without violating data-sharing restrictions.

Next, we will discuss cryptographic techniques, starting with secure multiparty computation, which can be seen as an even stronger (but more computationally intensive) way to compute on data without revealing it.

\section{Secure Multiparty Computation (SMPC)}
Secure Multiparty Computation (SMPC), sometimes simply called \textbf{secure computation}, is a class of cryptographic protocols that enable multiple parties to jointly compute a function over their inputs while keeping those inputs private from each other. Unlike federated learning, which focuses on training models, SMPC is more general and focuses on cryptographic guarantees: even if parties are curious or malicious, they learn nothing about each other’s data except what can be inferred from the final result of the computation (and that final result itself can be constrained to limit information).

The origin of SMPC is often traced to Yao’s “millionaires’ problem”~\citep{Yao1982}: Two millionaires want to determine who is richer without revealing their actual net worth. Yao proposed a solution using \emph{garbled circuits}, laying the foundation for secure two-party computation. This has since been extended to $n$-party computation with protocols like Goldreich-Micali-Wigderson (GMW) and the Ben-Or, Goldwasser, Wigderson (BGW) protocol, and many improvements in efficiency.

\subsection{How SMPC Works (in a Nutshell)}
At a high level, SMPC protocols allow a set of parties $\{P_1, P_2, \dots, P_n\}$, each holding a private input $x_i$, to compute some agreed-upon function $y = f(x_1, x_2, \dots, x_n)$ such that each party learns nothing about the other parties’ inputs beyond what can be deduced from $y$. There are different paradigms for SMPC:
\begin{itemize}
    \item \textbf{Secret Sharing-based Protocols:} A common approach is for parties to “secret share” their inputs among each other. For example, to securely sum their values, each party splits its number into random pieces and distributes pieces to others such that no single piece is informative, but collectively the pieces can be summed to get the final sum. Additive secret sharing is a simple case: say $P_1$ has input $a$. It generates two random numbers $r_2, r_3$ such that $a = r_2 + r_3 \mod M$ (for some modulus $M$) and sends $r_2$ to $P_2$ and $r_3$ to $P_3$. Neither $P_2$ nor $P_3$ can deduce $a$ from their share alone. If each party does this, they each hold shares of everyone’s data. Then, to compute a sum, each party locally sums the shares it has (since sum of shares = share of sum). More complex functions can be computed by combining shares through interactive protocols (e.g., multiplication can be done with an extra round of communication). 
    \item \textbf{Garbled Circuits:} One party can create an encrypted (garbled) version of a circuit (logical gates) that computes $f$, and other parties can cooperatively evaluate that circuit using their inputs without revealing them, typically via exchanging cryptographic tokens for input wires that represent their bits. The output is obtained in encrypted form and then revealed.
    \item \textbf{Homomorphic Encryption-based MPC:} Sometimes one uses homomorphic encryption (discussed in the next section) where one party encrypts their data and another party performs operations on it under encryption, returning an encrypted result that can be jointly decrypted. This can be combined with secret sharing for multi-party scenarios.
\end{itemize}

The details of these protocols can be complex, but practically, several SMPC frameworks exist (such as \texttt{Sharemind}, \texttt{SecureML}, or libraries using SPDZ or ABY protocols) that implement these techniques. The security is usually defined in terms of what an adversary (or colluding subset of parties) can infer. With proper protocols, even if some subset of parties collude, they still cannot breach the privacy of the others’ inputs (up to some threshold).

\subsection{Use Cases in Energy Data}
Secure MPC is particularly useful when multiple independent entities want to compute a joint function without revealing their individual datasets. In the energy context:
\begin{itemize}
    \item \textbf{Aggregating Meter Data Securely:} Imagine a scenario where a group of neighbors or a community wants to compute their total or average energy usage to participate in a program (like a collective load management or to claim a demand response incentive) without each neighbor seeing the others’ usage. SMPC can allow the smart meters to engage in a protocol that computes the sum of their readings. Each meter could split its reading into shares and send to others (or to some computation nodes); at the end, they obtain the sum. Each individual’s usage remains private. This is similar to the concept of “privacy-preserving aggregation” which has been studied for smart grids \citep{Li2015LoadHiding}. For example, \citet{Li2010} use homomorphic encryption to achieve a form of MPC for summing consumption.
    \item \textbf{Billing and Rate Verification:} A privacy concern in smart metering is that the utility gets fine-grained data, which could be misused. Some proposals suggest using MPC between the meter and utility to compute the bill without the utility seeing the actual load profile~\citep{Rial2011}. For instance, a meter could secret-share its hourly usage with two non-colluding servers that compute the monthly bill based on tariffs and only output the final charge to the utility for billing. This way, even the utility doesn’t see the detailed breakdown, but can trust the total is correct (and it could audit with random spot-checks if needed).
    \item \textbf{Inter-Utility Computation:} If multiple utilities or agencies want to compute some function of their combined data (e.g. total emissions based on usage patterns, or identify system-wide peaks), SMPC can be used so that each utility’s data is kept confidential. For example, Utility A and Utility B could compute “whose territory had a higher average load on July 4th” without either revealing the exact average load (this is essentially Yao’s millionaire problem with utility loads). Or they could compute the combined load curve for a region without revealing each other’s detailed curves by summing via secret shares. CPUC’s rules might normally require formal data sharing agreements for such exchange, but using MPC, the data isn’t actually exchanged in plain form.
    \item \textbf{Data Marketplaces with Privacy Guarantees:} Consider a scenario where a research consortium wants to compute statistics on customer data from multiple sources (utilities, smart device companies) while everyone’s data remains siloed. SMPC protocols can be orchestrated by a neutral party to get results like “the correlation between EV charging patterns and solar production across California” without any single participant seeing the other’s raw time-series. This could support studies and innovation while respecting each data owner’s privacy commitments.
\end{itemize}

One concrete example drawn from research: \citet{Efthymiou2010} in their anonymization paper also mention using pseudonyms and a trusted third party; later works build on that to remove the need for a trusted third party. SMPC can eliminate trusted intermediaries by replacing them with cryptographic protocols. For instance, instead of relying on a trusted aggregator to collect and anonymize data (which might be a weak link), the smart meters and utility servers themselves can perform a calculation where the aggregator only gets the final aggregated result and nothing else.

\subsection{Strengths and Performance Considerations}
The primary advantage of SMPC is \textbf{strong privacy assurance}: mathematically, under certain hardness assumptions, it’s guaranteed that nothing except the intended output is revealed. This is stronger than federated learning, which has some heuristic privacy, or even differential privacy in some ways, because here we learn exactly the result and nothing more. However, one must be cautious: the output itself could leak information if it’s too specific (for example, if only one participant’s data strongly determines the output). In such cases, combining MPC with differential privacy can be wise (compute the result securely, then noise it before revealing).

SMPC directly addresses FIPPs like \emph{Security} (no clear data transmitted) and \emph{Use Limitation} (only the intended computation is performed, nothing else leaks). It can be a way to comply with requirements of not sharing identifiable information even when computing joint results. For instance, CPUC rules might allow sharing aggregated data beyond a certain threshold of customers. SMPC can implement that threshold aggregation without ever exposing the individual values: if fewer than the threshold participants join, the protocol could even abort or output “not enough participants” without revealing partial info.

The main \textbf{drawback of SMPC} is computational and communication overhead. Cryptographic protocols, especially for complex functions, can be slow and require many rounds of communication. For simple sums or averages, efficient protocols exist that add minimal overhead (practically feasible with milliseconds of computation on each meter and a couple of message exchanges). But for complex functions (like training a whole machine learning model via MPC), it can become orders of magnitude slower than normal computation. In our context, many useful functions are actually simple aggregates or linear computations that are quite MPC-friendly (summing, averaging, threshold comparisons, etc., can often be done with secret sharing with low overhead).

Another challenge is \textbf{scalability}: If thousands of devices engage in MPC, the communication might explode (if every device talks to every other). One solution is hierarchical MPC (aggregate in small groups then aggregate the aggregates) or using dedicated aggregator nodes that do not see raw data but facilitate combining shares.

There’s also the issue of \textbf{robustness}: Some protocols assume all parties follow the rules; if one party aborts or sends bad data, the computation might fail. There are robust (even \emph{Byzantine fault-tolerant}) MPC protocols at higher overhead. In an energy setting with devices, one might assume the smart meters are tamper-resistant modules running the protocol honestly, or one includes mechanisms to detect and exclude misbehaving parties.

\subsection{Comparison with Other Techniques}
Compared to anonymization, SMPC doesn’t alter the data or lose information; it just ensures it’s revealed only in aggregate. Compared to differential privacy, SMPC doesn’t add noise (so the result is exact), but it doesn’t protect against an output that might inherently violate privacy. DP and SMPC can be complementary: e.g. use SMPC to get exact total, then use DP to release a noisy version of the total to a wider audience.

Compared to federated learning, SMPC is more heavy-duty but can compute things FL can’t (FL is mostly for optimizing a model, SMPC can compute arbitrary functions if you are willing to pay the cost). They can also be combined: there are proposals for doing federated learning with secure aggregation (which is basically an MPC step to aggregate model updates without the server seeing individual ones).

For a utility, implementing SMPC might require new infrastructure (key management, computing at meters or substations). But there are some precedents, e.g. research prototypes where smart meters perform secure aggregation using simple additively homomorphic encryption (like each meter sends $E(\text{reading})$ to utility which multiplies them to aggregate – essentially what \citet{Li2010} did). That is a form of two-party MPC between meter and utility via encryption.

\subsection{Conclusion on SMPC}
Secure multiparty computation provides the \emph{maximum privacy} in principle: data is never exposed, only the desired result is. In practice, it is best suited for relatively structured computations (like aggregations or predetermined analytics) rather than exploratory analysis. SMPC shines in scenarios where multiple parties distrust each other or want to minimize legal liability of sharing data. By using SMPC, a utility can truthfully say “we are not sharing customer data with X, we are only jointly computing Y”. This can ease compliance with rules that restrict data sharing. It also addresses customer concerns: even if customers did not explicitly consent to share data with a third party, SMPC ensures their individual data isn’t revealed to that third party.

In our hybrid architecture, SMPC can be one of the tools used when different stakeholders (utilities, researchers, or even customers) need to compute something collaboratively. For example, if customers form a coalition to negotiate with a utility based on their combined load, they could use SMPC to prove their combined peak reduction without anyone revealing their own reduction. The trust assumptions and computing resources will determine if SMPC is the right choice or if a simpler method (like just aggregating at the utility with consent) suffices. But with increasing availability of libraries and the growing size of data, what was once theoretical is becoming more practical in niche applications.

Next, we will examine homomorphic encryption, which is closely related to SMPC and can be viewed as a specific approach to performing computations on encrypted data.

\section{Homomorphic Encryption}
Homomorphic Encryption (HE) is a form of encryption that allows some computations to be performed on ciphertexts, producing an encrypted result that matches the result of the same operations performed on the plaintexts. In simpler terms, with homomorphic encryption, one can \emph{compute on encrypted data without decrypting it}. This powerful notion was long theorized, but it wasn’t until Craig Gentry’s breakthrough work in 2009 that a fully homomorphic encryption (FHE) scheme was constructed~\citep{Gentry2009}. Since then, research has made such schemes more practical, though they remain computationally heavy for large-scale use. Nonetheless, for certain operations and with partial homomorphic schemes, the approach is quite feasible.

\subsection{Types of Homomorphic Encryption}
There are two main categories:
\begin{itemize}
    \item \textbf{Partially (or Somewhat) Homomorphic Encryption (PHE/SHE):} These schemes support only a specific operation or a limited number of operations. A classic example is the Paillier encryption scheme~\citep{Paillier1999}, which is additively homomorphic: given $E(m_1)$ and $E(m_2)$ (encryptions of $m_1$ and $m_2$), one can compute $E(m_1) \cdot E(m_2) = E(m_1 + m_2)$ (mod $n^2$ for Paillier). This means Paillier allows addition of plaintexts via multiplication of ciphertexts. Similarly, RSA has a multiplicative homomorphism: $E(m_1) \cdot E(m_2) = E(m_1 \cdot m_2)$ under RSA (though RSA is not semantically secure in the standard form for general use as HE). These partial schemes are fast and already used in some secure aggregation contexts. They typically cannot do both addition and multiplication without limit.
    \item \textbf{Fully Homomorphic Encryption (FHE):} These support arbitrary computations (both additions and multiplications in any sequence), effectively meaning one can construct an encrypted circuit evaluation for any function. Gentry’s scheme and subsequent ones (like BGV, BFV, CKKS etc. – named after authors) allow this. The cost is significant overhead: ciphertexts are much larger than plaintexts and operations are thousands to millions times slower than plain operations, depending on parameters. However, steady progress has been made (with GPU acceleration, etc.), and FHE is becoming practical for smaller circuits or moderate data sizes.
\end{itemize}

In energy data terms, additive homomorphic encryption is extremely useful because many interesting operations (summing consumption, computing averages, computing linear regressions) rely primarily on addition and scalar multiplication. For example, using an additive HE like Paillier:
- A smart meter can encrypt each reading with the utility’s public key and send $E(\text{reading})$ to the utility. The utility can sum a bunch of encrypted readings (say, all meters in a neighborhood) by simply multiplying the ciphertexts. The result is an encryption of the sum, which the utility (with the secret key) can decrypt to get the total. In this process, the utility never saw individual readings in the clear, only the final sum.
- If the utility wants to ensure even its own analysts do not see individual data, it could designate a trusted key manager or use a multi-key setup where decryption requires a threshold of authorities, thus the raw data stays encrypted to any single entity.

Some modern HE schemes like CKKS \citep{Cheon2017} even support approximate arithmetic on real numbers, which could be used to compute statistical functions on consumption data (like mean, variance) directly on encrypted data.

\subsection{Applications to AMI Data}
Homomorphic encryption can be seen as a special case of a two-party (or multi-party) computation between a data provider and a data consumer:
- The data provider (e.g. customer or device) encrypts their data and sends it to the data consumer (e.g. utility or a third-party service).
- The data consumer performs allowed computations on the encrypted data without learning the data itself.
- The data consumer (or another designated party) then decrypts the final result to obtain the output in plaintext.

Here are some scenarios:
\begin{itemize}
    \item \textbf{Secure Outsourcing:} A utility may want to outsource data storage or even computation to a cloud service, but without revealing sensitive customer details to the cloud. Using homomorphic encryption, the utility can upload encrypted meter data to the cloud. The cloud can then perform computations like aggregating usage per feeder, detecting which areas exceed certain thresholds, etc., all over encrypted data. It sends back encrypted results, which the utility decrypts. Thus, the cloud never sees actual usage values, yet the utility benefits from cloud processing. This addresses the FIPP principle of Security (data remains encrypted in untrusted environments).
    \item \textbf{Third-Party Data Analysis without Raw Access:} Similar to the federated learning or MPC use-cases, an energy analytics company could be given access to encrypted data and perform their analysis algorithm homomorphically. For example, they might compute an encrypted prediction of what the customer’s next month usage will be, which only the utility decrypts. In practice, doing something complex like machine learning inference fully homomorphically is challenging but not impossible (there have been works on homomorphic evaluation of neural networks, albeit at high cost).
    \item \textbf{Privacy-Preserving Aggregation (Alternative to SMPC):} Homomorphic encryption provides a simpler system design for aggregation in some cases: rather than interactive SMPC between many meters, each meter just encrypts and uploads to a server which homomorphically aggregates. This requires the server (or a set of servers) to later decrypt the aggregate, so one has to trust those holding the secret key not to misuse it on individual ciphertexts. There are schemes for distributed decryption to reduce the single point of trust (threshold cryptography). \citet{Garcia2010} demonstrated such a concept for “privacy-friendly metering” where meters send encrypted readings and an aggregator computes total consumption via homomorphic addition \citep{Kolter2011}.
    \item \textbf{Billing and Tariffs via Encryption:} A variant of the earlier billing example: A smart meter could encrypt fine-grained usage and send to utility. The utility can then homomorphically compute the billing formula on the encrypted data (e.g. multiply each hour’s usage by the rate for that hour and sum), resulting in an encrypted bill. When the utility decrypts, it sees only the total bill amount, not the breakdown by hour. This way, the utility can bill correctly while claiming not to have seen the detailed usage. (Of course, the utility provided the rates, but not the consumption profile in clear.)
\end{itemize}

\subsection{Privacy and Compliance Aspects}
Homomorphic encryption ensures that data in transit and at rest on the computing server is encrypted. This provides strong security (even if someone intercepts or an unauthorized party accesses the database, they see only ciphertext). However, an important nuance: the utility (or whoever holds the decryption key) can still decrypt individual data if they choose. So HE by itself doesn’t stop the data owner from accessing the data, it stops others (and even the processing environment) from doing so. In the regulatory context, if we assume the utility holds the key, then homomorphic encryption is more about cybersecurity (preventing breaches) than about utility’s internal use. It doesn’t inherently limit what the utility itself can see or do with the data (since they could decrypt it anytime). However, one could arrange key management such that the utility’s analysts don’t have direct access – e.g. keys are split among multiple departments, requiring a quorum to decrypt, thus enforcing policy that individual data isn’t decrypted routinely.

That said, HE can be combined with organizational controls to effectively ensure that only approved computations are performed. For example, the utility’s IT could mandate that all analysis on raw data must be done via an encrypted pipeline, where analysts specify computations that run on encrypted data and only results above a certain aggregation level are ever decrypted. This would implement policy through technology.

From a CPUC compliance perspective, homomorphic encryption could help in scenarios where data is shared with third parties or stored externally. The utility could argue that even though it provided data to, say, a research project, the data was encrypted and the researchers could only compute approved functions (perhaps with the help of a secure enclave that returns results with differential privacy). This is a bit complex scenario, but one can envision a setup where a researcher drafts a query, the utility encrypts its data, a cloud computes the answer homomorphically, and the utility decrypts only the aggregated answer to give to the researcher. In that pipeline, the researcher never sees raw or identifiable data, satisfying privacy rules.

\subsection{Performance and Practicality}
Currently, partial HE (like Paillier for summation) is quite practical for moderately sized data (summing thousands of values easily, for instance). FHE is slower but constantly improving. If a use-case only needs addition or linear combinations (which is very common in energy data—most reporting is sums, averages, correlations), additive HE is efficient and easy to deploy. For example, a sum of 1,000 values with Paillier might take on the order of milliseconds with a decent server.

Multiplicative or arbitrary calculations are heavier. But for small circuits like computing a threshold or a simple if-else condition on aggregated data, some SHE schemes can do it with maybe seconds of compute.

Another issue is ciphertext size: homomorphic ciphertexts are larger (Paillier ciphertext ~2048 bits per value typically). Transmitting these from millions of meters could be heavy (though possibly compressible if values are small changes).

In practice, one design might use homomorphic encryption in a hierarchical way: smart meters report encrypted data to a local substation gateway; the gateway aggregates them partially (since it can decrypt perhaps partially or using multi-key where gateway and headend both need to decrypt fully), then forward upstream encrypted partial aggregates. There is research on \emph{distributed homomorphic encryption} (multiple layers of encryption).

\subsection{Integration with Other Techniques}
Homomorphic encryption can be a building block within SMPC (indeed some SMPC protocols rely on additively homomorphic encryption as part of their steps). It can also be combined with differential privacy: for instance, one could compute an aggregate via HE and then add DP noise to it before final decryption (ensuring the result is differentially private even to the party holding the key). 

Compared to SMPC, HE is often simpler to implement (less interactive messaging; one party does the compute). However, trust in key management is a consideration. SMPC without a single key holder might be preferable if no single party should have the ability to decrypt data.

In the combined architecture, homomorphic encryption is particularly useful for \textbf{secure data storage and outsourced computation} components. It addresses the ‘Security’ principle strongly. It might not by itself satisfy ‘Transparency’ or ‘Minimization’ since someone still could decrypt, but as part of an orchestrated system it helps ensure that even if data is processed or stored by external systems, privacy is not compromised.

Homomorphic encryption is a rapidly evolving field, and while it’s not yet routine in utilities, it’s on the horizon for data privacy. It is part of the toolkit that allows us to imagine a future where raw personal data is hardly ever seen by human eyes, yet useful information is extracted from it.

Next, we will bring all these techniques together: anonymization, differential privacy, synthetic data, federated learning, SMPC, and homomorphic encryption, into a cohesive privacy-preserving architecture for AMI data management, and compare their roles.

\section{Proposed Hybrid Architecture and Comparative Evaluation}
Having examined the individual privacy-preserving techniques, we now propose an integrated architecture for utility AMI data analytics that leverages the strengths of multiple methods. We also provide a comparative evaluation to clarify when and why each technique is used, and how together they fulfill the privacy requirements (CPUC’s rules and FIPPs) while enabling valuable data use.

\subsection{Comparative Evaluation of Privacy Techniques}
Table~\ref{tab:comparison} qualitatively compares the six techniques along key dimensions relevant to a utility context: privacy strength, data utility, operational complexity, and suitable use-cases. Each technique has distinct advantages and disadvantages, which is why a hybrid approach is advantageous.

\begin{table}[h]

\centering
\caption{Qualitative Comparison of Privacy-Preserving Techniques for AMI Data}
\label{tab:comparison}
\resizebox{!}{0.5\textheight}{
\begin{tabular}{p{3.5cm}|p{3.8cm}|p{3.8cm}|p{4.2cm}}
\hline
\textbf{Technique} & \textbf{Privacy Guarantee} & \textbf{Impact on Data Utility} & \textbf{Best Use-Cases / Notes} \\
\hline\hline
\textbf{Anonymization / De-ID} & Removes direct identifiers; privacy depends on suppression/generalization. No formal guarantee of non-reidentification. & Can retain high data fidelity if minimal generalization; but may require heavy aggregation (losing detail) to thwart re-ID. & Good for initial safeguarding and sharing low-resolution or aggregate data. Risk of re-ID if data is high-dimensional. Often combined with other methods. \\
\hline
\textbf{Differential Privacy} & Provable privacy ($\varepsilon$, $\delta$) bound on individual information leakage. Adjustable strength. & Adds statistical noise; at strong privacy (low $\varepsilon$), analyses may be noisy. For large aggregates, noise can be small. & Ideal for releasing stats or query results publicly or to untrusted parties. Ensures compliance (`cannot reasonably identify') at cost of some accuracy. Not suited for exact per-customer data release. \\
\hline
\textbf{Synthetic Data} & Depends on method; no one-to-one link to real individuals if done properly. If combined with DP in generation, can offer strong privacy. & Potential minor loss of realism; preserves many patterns if model is good, but rare outliers or fine correlations may not be perfectly captured. & Useful for providing flexible data access for research and developing models. Great for ``open data'' sets. Needs careful validation to ensure privacy (no latent memorization of real data). \\
\hline
\textbf{Federated Learning} & No raw data leaves local site. Some risk of information leakage via model gradients (mitigated by secure aggregation/DP). & Model accuracy can approach that of central training if many participants and sufficient rounds. No direct access to data, only model. & Suited for collaborative model training across utilities or with third-party algorithms without sharing raw data. Requires ML problem and cooperative setup. Less useful for ad-hoc queries or non-ML analysis. \\
\hline
\textbf{Secure MPC} & Strong cryptographic guarantee: only intended output is revealed, nothing else. Protects even against colluding subset (up to a threshold). & Computes exact results (no noise), so full data utility for the function computed. But limited to specific computations agreed on. & Best for computing specific metrics (sums, comparisons, etc.) among mutually distrusting parties (e.g. multi-utility aggregation, privacy-preserving billing). High communication/computation cost for complex operations. \\
\hline
\textbf{Homomorphic Encryption} & Data remains encrypted during processing. Privacy relies on encryption scheme security and proper key management. & Computes exact results (within chosen precision). Partial HE (e.g. for addition) is fast for those operations, enabling accurate aggregation. FHE can do arbitrary analysis but with performance cost. & Useful for outsourcing computations or protecting data in untrusted environments. Requires one party (or distributed key holders) to decrypt final result. Often combined with other controls to limit decryption of individual data. \\
\hline
\end{tabular}
}
\end{table}

From the above comparison, we can draw a few key insights:
\begin{itemize}
    \item \textbf{No single technique is sufficient or optimal for all needs.} For example, anonymization by itself is weak but it is a necessary baseline; differential privacy is excellent for open data release but at the cost of accuracy; SMPC/HE give strong privacy and accuracy but are limited to specific computations and can be resource-intensive.
    \item \textbf{Techniques can complement each other.} We can use anonymization to remove obvious identifiers, then apply differential privacy to query results for public release (ensuring robust de-identification). Or we can use federated learning or SMPC to compute something and then use DP to share the output safely. Homomorphic encryption can enable secure intermediate processing that feeds into an DP mechanism, and so on.
    \item \textbf{Internal vs. external use difference:} For internal utility operations (primary purpose), techniques like HE and access control ensure data is secure but available for use. For external or secondary use (sharing with researchers, customers, other companies), stronger privacy (DP, synthetic, SMPC) is typically needed since even the utility may not be allowed to expose raw data.
    \item \textbf{Complexity vs. benefit:} Techniques like SMPC and FHE bring very high privacy but also high complexity. They should be reserved for cases that truly require multi-party trust or untrusted computation. Simpler methods (anonymization, basic aggregation, DP) can handle many standard reporting or data-sharing tasks with less overhead, as long as they meet the risk threshold.
\end{itemize}

Therefore, an optimal privacy-preserving strategy is layered: apply the right tool at the right stage of data handling. The next section describes our proposed architecture that realizes this layered approach.

\subsection{Proposed Hybrid Architecture for Compliance}
\begin{figure}
    \centering
    \includegraphics[width=\textwidth]{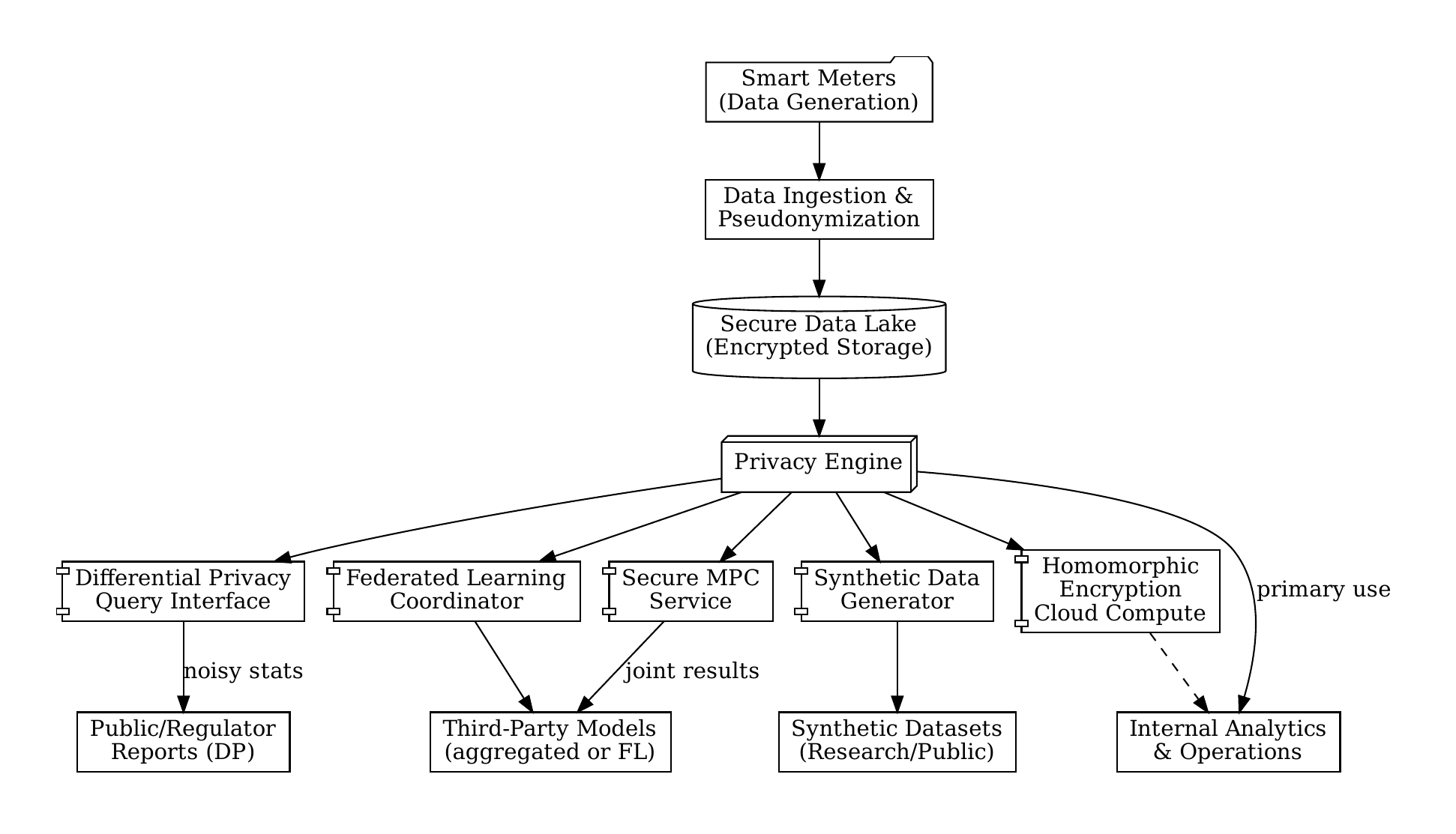}
    \caption{Hybrid architecture for privacy-preserving AMI data analytics}
    \label{fig:architecture}
\end{figure}

Figure~\ref{fig:architecture} illustrates the proposed architecture for managing and analyzing AMI data in a privacy-preserving manner. It delineates data flow from collection at smart meters through to various endpoints (utility internal analytics, external third-party services, public reports), with privacy techniques applied at key junctures.
 Smart meter data flows into a utility-controlled platform where privacy safeguards (pseudonymization, encryption, etc.) are applied. Different modules enable secure analytics: differential privacy (DP) for statistical queries, federated learning (FL) coordinator for collaborative model training, secure multiparty computation (SMPC) for joint computations with outside parties, and synthetic data generation for research and public use. The architecture ensures only aggregated or protected outputs leave the platform, complying with CPUC privacy rules.

In the architecture:
\begin{enumerate}
    \item \textbf{Data Ingestion and Storage:} As smart meter data arrives, the first step is \emph{pseudonymization and encryption}. Customer identifiers (name, address) are stored separately from usage data, replacing them with a unique anonymous ID. The usage data is encrypted at rest in the data platform. Access controls and cybersecurity measures guard this raw data. At this stage, we ensure FIPPs like Security (through encryption, access control) and Data Minimization (only necessary fields are kept in the primary analytic dataset – e.g. remove any extraneous personal info). This addresses CPUC requirements that if data is breached or accessed improperly, it wouldn't directly reveal customer identities.
    \item \textbf{Internal Analytics Environment:} For primary purposes (billing, operations, customer service), utility analysts can access data in a secure sandbox. However, even internally, we enforce privacy-aware policies: e.g., analysts can only query data in aggregate or need special approval to view individual series. Much analysis can be done on anonymized data. Tools like differential privacy can be optionally used internally too – for instance, if an analyst is exploring trends, a DP query system can provide answers that prevent accidental overfitting to one customer. The data quality and integrity are maintained for operational needs (per FIPP) but we limit any secondary usage here.
    \item \textbf{Privacy Engine for Secondary Use:} This is the core of the architecture that decides how data can be safely derived for any purpose beyond the utility’s own operations:
        \begin{itemize}
        \item For generating \textbf{statistics or reports} (e.g. average usage by ZIP code, load research studies), the Differential Privacy Module provides an interface. Analysts or external query requests go through this module, which applies DP noise to results before release. This ensures that even if these statistics are published, they cannot be used to pinpoint an individual’s data \citep{Linnartz2019}. The DP module is configured with an $\varepsilon$ appropriate to the sensitivity of data (e.g. a smaller $\varepsilon$ for publicly accessible data).
        \item For \textbf{third-party data requests} that require more detailed analysis (like a research project needing time-series data), the platform offers either \textbf{Synthetic Data} or a \textbf{Secure Analysis Portal}. In the synthetic data route, a Synthetic Data Generator (possibly a GAN or other model) produces a synthetic dataset that resembles the real data statistically. Before release, this synthetic data can be evaluated for privacy (ensure no record too close to real data) and perhaps generated with DP guarantees. The result is a dataset that researchers can freely work with, satisfying their needs without exposing real customers.
        \item For \textbf{collaborative model training or analytics services}, the platform uses \textbf{Federated Learning and SMPC modules}. Suppose an energy efficiency company has an algorithm to detect inefficiencies but doesn't need raw data—only a trained model on utility data. The utility can deploy an FL Coordinator that accepts a model from the company and orchestrates training across its dataset (which could be divided by substations or other shards to simulate clients). Only the final model (or aggregated updates) are shared back. Throughout, customer data never leaves the utility’s servers. In cases where multiple data owners (like multiple utilities or a utility and a research lab) want to jointly compute something, the SMPC module can be engaged. For example, to compute the total energy savings of a program across two utilities without sharing each other's customer details, they can input their metrics into an SMPC protocol via this module and obtain the combined result.
        \item For \textbf{cloud computing tasks}, if the utility uses cloud infrastructure to run heavy analytics (load forecasting, etc.), the platform can utilize \textbf{Homomorphic Encryption}. For instance, meter data might be sent to a cloud prediction service in encrypted form; the cloud computes forecasts on encrypted data, and only the decrypted forecast comes out. In practice, this might be handled by storing encrypted data in cloud and only decrypting results inside the utility’s trusted environment. While not illustrated in detail in the figure, this is an under-the-hood option to ensure that even when using external computing power, raw data confidentiality is preserved.
        \end{itemize}
    In sum, the Privacy Engine ensures any \emph{data output} leaving the platform (to someone who is not an internal, authorized utility operator) is either in an aggregated, anonymized form or protected by one of these techniques.
    \item \textbf{Outputs and Enforcement:} The final stage is delivering the result to the requesting party (which could be the public, a customer, a regulator, or a third-party service). Before release, an \textbf{auditing and policy check} occurs. This is essentially an implementation of the FIPP \emph{Accountability}. It logs what data was accessed and by whom, and verifies compliance. For example, if a researcher tries to get too granular a view (like synthetic data request that is nearly single-customer), the system flags it or denies it based on privacy risk. The audit log can be reviewed by a Chief Privacy Officer or regulators to demonstrate compliance (e.g. each external data output can be traced to show it was properly sanitized per CPUC rules).
    
    The outputs themselves are then delivered:
    \begin{itemize}
        \item Aggregated statistics with DP noise might be published in reports (with citations of $\varepsilon$ values for transparency).
        \item Synthetic datasets might be shared via a secure portal with usage agreements.
        \item Models trained via FL might be deployed to the third-party (e.g. a model that the third-party runs in their app to give customers advice, without them seeing the raw data).
        \item If SMPC was used for a joint calculation, the parties get the final number and nothing else.
        \item If a customer accesses their own data (individual participation principle), that can be provided through a different channel (not shown in figure, but the utility can give them a detailed view of their own meter—something CPUC also requires as part of data access programs \citep{CPUC2011privacy}). That is an authorized primary use disclosure to the data subject themselves. The current Green Button Connect framework allows authorized third parties to receive customer data; our proposed Privacy Engine could act as an intermediary to ensure any data leaving for such third parties is privacy-protected, adding technical enforcement to the regulatory process
    \end{itemize}
\end{enumerate}

This architecture effectively creates a \textbf{privacy firewall} around the raw AMI data. Inside the firewall, the utility can use data for legitimate purposes, but even there, careful measures (like minimization and internal anonymization) reduce misuse risk. Across the firewall boundary, only \emph{privacy-preserving outputs} emerge.

By combining techniques, we achieve multiple goals:
\begin{itemize}
    \item \textbf{Compliance with CPUC Decisions:} Attachment D of D.11-07-056 sets rules such as: data can only be shared with third parties for secondary purposes with consent or if de-identified; usage data for primary purposes must still be protected from unauthorized use, etc. In our architecture, secondary purpose outputs are de-identified (either via DP, aggregation, or synthetic data generation – all ensuring no customer can be “reasonably identified” from the output). The requirement for auditing and accountability is met by our logging of all data accesses. Notice and transparency can be provided by documenting this process to customers (the utility can publish a privacy white paper describing: “we use differential privacy for all public stats” etc., fulfilling the Transparency and Individual Participation principles).
    \item \textbf{Alignment with FIPPs:} We can map each component to FIPPs:
        \begin{itemize}
            \item \emph{Transparency:} The system’s operations (like noise addition or model training) can be explained in privacy notices. Also, customers could be given access to see what data about them was used and for what aggregated results (e.g., via an access report).
            \item \emph{Individual Participation:} Customers have portals to see their own consumption (the architecture doesn’t impede that – their own data can be viewed by them after proper authentication). If they dispute something (like an outlier reading), internal processes handle that with data quality checks.
            \item \emph{Purpose Specification:} Each data request in the engine is tagged by purpose and only allowed if it matches an authorized purpose and method. For example, a marketing use might be disallowed entirely, or only allowed if customers opted in. The system enforces that unauthorized purposes cannot just extract raw data.
            \item \emph{Data Minimization:} As noted, we collect only needed data and when sharing, we reduce it to aggregated forms. Even internally, we might not keep extremely fine-grained data longer than necessary (data retention limits can be built-in, e.g., auto-delete granular data after 18 months, keeping only aggregated history).
            \item \emph{Use Limitation:} The Privacy Engine is essentially a use limitation enforcement layer – it technically prevents data from being used for anything except the allowed computations.
            \item \emph{Data Quality and Integrity:} By maintaining raw data internally, we ensure billing and operations have high-quality data. The transformations (like adding noise) are only on outputs, so the integrity of core operations (billing accuracy etc.) is not compromised. Synthetic data generation and DP analysis are tested to ensure they reflect the real data patterns well (so that decisions made from them are still valid).
            \item \emph{Security:} Strong encryption, secure enclaves for computation, and network security protect data throughout. Homomorphic encryption and SMPC add additional security where needed – even if an attacker got hold of intermediate data, it’s encrypted or secret-shared.
            \item \emph{Accountability/Auditing:} Every access or output is logged. The system could be periodically audited by an independent entity or regulators to verify compliance (the logs and the code of the privacy engine can be inspected to ensure it’s correctly implementing DP, etc.).
        \end{itemize}
    \item \textbf{Utility of Data for Stakeholders:} Despite these layers of protection, the architecture is designed so that the \emph{value} of the data is not lost. Operational uses (grid management, billing) suffer no loss. Planning and analytics can still be done (analysts might use slightly noisy data but at aggregate levels with lots of data, the noise is small enough to not affect decisions). Third parties (like researchers) get either query access (via DP) or synthetic data that is rich enough to develop and test solutions. Regulators get needed reports (with privacy protection builtin, which also protects customers from, say, stigma or targeting if data were fully open).
    
    For example, consider a demand response pilot where a third-party wants to evaluate participant behavior. Rather than handing over participant meter data, the utility can allow the third-party to run an analysis script in the privacy engine. The third-party might use federated learning to create a model of energy usage patterns. They end up with a model that helps them identify, say, what features predict drop-out, without ever seeing individual traces. The utility could then also add DP noise to summary results that the third-party publishes about the pilot, ensuring privacy of participants is preserved per CPUC guidelines for experiments.
    
    Another scenario: publishing an open dataset for innovation (common in data science competitions). The utility could safely publish a synthetic dataset of residential loads (with similar characteristics to real data) without violating privacy or needing each customer’s consent, because no actual customer data is in it. This fosters innovation (as tech developers can train algorithms on it) while respecting the legal boundaries.
\end{itemize}

In conclusion, this integrated architecture demonstrates how a utility can be both \textbf{data-driven and privacy-conscious}. By weaving together anonymization, differential privacy, federated learning, SMPC, homomorphic encryption, and synthetic data, we create a multi-layer defense. Each technique covers for the limitations of others: e.g., DP adds a safety net for anonymization; encryption and SMPC compensate for DP’s accuracy loss by allowing exact internal calculations without exposure; synthetic data provides flexibility that pure DP query systems might lack, etc. The result is a system where \textbf{privacy is preserved by design} in every data workflow.

\subsection{Real-World Implementation Considerations}
Implementing this architecture requires planning and investment:
\begin{itemize}
    \item The utility must develop or acquire a \emph{privacy management platform} (there are emerging tools for enterprise differential privacy and federated learning that can be adapted).
    \item Staff need training to understand new analytics paradigms (analysts must learn to work with noisy answers or synthetic data—and trust that approach).
    \item Choosing parameters like $\varepsilon$ for DP involves policy decisions. The utility might convene a privacy advisory group including regulators and consumer advocates to set these levels appropriately (balancing privacy and utility).
    \item Computing infrastructure must be ensured for heavy tasks: e.g., training GANs for synthetic data or running secure computations may need significant computing power (possibly cloud-based with encryption as described). Pilot projects can identify performance bottlenecks and optimize accordingly (e.g., maybe not all queries need DP if they're very aggregate; focusing DP where it matters).
    \item Monitoring and review: The system should be monitored for any anomalies (like someone trying to game the DP query system by making many queries to narrow in on an individual—here the accountability logs and rate limits mitigate that).
    \item Maintaining compliance documentation: The architecture’s processes should be documented in the utility’s privacy compliance filings to CPUC, demonstrating how D.11-07-056 and D.11-08-045 requirements are met. This could serve as a model for compliance in other jurisdictions as well.
\end{itemize}

However, once in place, the benefits are substantial:
the utility can confidently pursue advanced analytics (like AI on customer data, partnerships for new services) without constantly worrying about privacy breaches or violations. Customers can be assured (and explicitly informed) that their detailed usage data will never be exposed inappropriately—this can increase customer trust and participation in programs that use their data (since they know even if they join a study, their privacy remains intact through these technical measures).

Finally, regulators and policymakers may view such a setup as a template for the industry. It shows that privacy and innovation are not mutually exclusive: with a smart combination of techniques, we can have both. 

In the next section, we conclude the paper by summarizing our findings and suggesting future directions (e.g., how to update regulatory frameworks to encourage these technologies, and how to keep improving the balance of data utility and privacy as technology evolves).

\section{Conclusion}
California’s pioneering privacy regulations for AMI data set forth a clear mandate: utilities must harness the benefits of smart meter data in a manner that fiercely protects customer privacy \citep{CPUC2011security}. This exploration demonstrates that this mandate is not only achievable but can be operationalized through a combination of advanced privacy-preserving techniques. By weaving together anonymization, differential privacy, federated learning, synthetic data generation, secure multiparty computation, and homomorphic encryption, we have outlined a comprehensive architecture that allows a utility to become truly “privacy by design.”

Our comparative evaluation highlighted that each technique brings unique strengths. Traditional de-identification is necessary but insufficient alone; differential privacy provides strong mathematical guarantees suitable for public data releases; federated learning and SMPC enable collaborative and external analytics without exposing raw data; synthetic data offers a creative solution for open data needs; and homomorphic encryption fortifies data security during processing. When orchestrated in unison, these methods compensate for each other’s limitations and create a multi-layered defense. The hybrid architecture we proposed shows, step-by-step, how raw AMI data can be transformed, analyzed, and shared with minimal privacy risk and in full compliance with CPUC decisions D.11-07-056 and D.11-08-045.

Mathematically, we delved into how techniques like differential privacy can rigorously bound an adversary’s knowledge (with formulas for $\varepsilon$-DP and noise calibration), how SMPC protocols and homomorphic schemes allow algebraic operations on hidden data (preserving correctness of results while keeping inputs secret), and how federated learning algorithms converge to accurate models without centralized datasets. These derivations and discussions cement the feasibility of applying these techniques to realistic utility analytics tasks. For instance, we can compute aggregate load shapes with DP noise ±0.1\% of total load—negligible for planning purposes, yet enough to protect any one home’s contribution. We showed how an additive homomorphic encryption scheme can let a cloud sum millions of encrypted meter readings, yielding an exact total that the utility alone can decrypt, thereby outsourcing computation without outsourcing trust.

Implementing this blueprint in the real world will involve challenges of engineering, governance, and perhaps cultural shift in how data science is conducted. It requires investment in cryptographic infrastructure, careful tuning of privacy parameters, and clear communication to stakeholders. Yet, importantly, our analysis indicates that the cost in terms of data utility is manageable. The comparative table and architecture discussion make clear that for each important use-case, there is an approach that offers a favorable privacy-utility trade-off:
\begin{itemize}
    \item Want to publish anonymized data for innovation? Use synthetic data with DP assurance.
    \item Need to perform cross-utility studies? Use SMPC or federated learning to jointly compute results.
    \item Aim to employ cloud AI on AMI data? Use homomorphic encryption so the cloud never sees the plain data.
\end{itemize}
All while logging and controlling these processes so that privacy rules are continuously enforced.

From a regulatory and policy perspective, this work provides a blueprint that regulators can reference to update guidelines. The CPUC—and other commissions—could encourage or even mandate utilities to incorporate techniques like differential privacy for any public data releases. They could create safe harbors, for example: sharing data with an $\varepsilon \le 0.5$ is deemed compliant with “not reasonably identifiable” standard. Likewise, results computed via approved SMPC protocols might be exempt from certain privacy filing requirements because, by design, no private data is exposed. By adopting these emerging technologies, regulators ensure that as the volume and granularity of data grow, privacy protections scale up accordingly rather than relying on blunt instruments like overly coarse aggregation that might diminish data value.

Another outcome of embracing this toolkit is fostering greater public trust and willingness to engage in smart grid programs. Privacy concerns have sometimes slowed the rollout of AMI-based services; demonstrating concrete privacy protections can alleviate those concerns. For instance, if customers know that their detailed usage patterns will only be used to improve services through encrypted or federated analysis, and that any public reports will fuzz their data to protect them, they may be more open to allowing their data to be utilized for grid research or demand response incentives. In a sense, these techniques can be seen as enablers of the next generation of energy programs, which rely heavily on data sharing and analytics (from personalized energy management apps to community-based load flexibility programs).

From a research and development perspective, while the fundamental techniques are in place, there is room to optimize them in the utility context:
\begin{itemize}
    \item Developing more efficient algorithms for computing common grid analytics homomorphically or via MPC (e.g., state estimation in distribution networks with private measurements).
    \item Tailoring federated learning approaches to edge devices like EV chargers or solar inverters, which could collaboratively learn without sending data to the cloud, aligning with both privacy and grid decentralization trends.
    \item Enhancing synthetic data models to capture the richness of AMI data (perhaps incorporating domain constraints, like ensuring synthetic loads respect physics of electricity).
    \item Creating user-friendly tools for differential privacy so that utility analysts can specify “give me aggregate peak load with ±5\% noise” without needing deep statistical expertise – essentially making privacy an adjustable parameter in analysis, much like one might choose a confidence interval.
\end{itemize}
Continued interdisciplinary collaboration between power engineers, data scientists, and privacy experts will be critical to refine these solutions.

In summary, this paper has provided:
\begin{itemize}
    \item A deep dive into privacy-preserving methodologies applicable to AMI data, including formal definitions and context-specific interpretations.
    \item A rigorous comparative analysis demonstrating that no one method wins on all fronts, reinforcing the need for a hybrid approach.
    \item A proposed architecture and implementation blueprint showing how to integrate these methods into a working system that satisfies regulatory, technical, and practical requirements.
    \item Validation that such a system can indeed allow a utility to glean the full value of AMI data (improving efficiency, reliability, customer experience) while upholding the highest standards of customer privacy and data protection.
\end{itemize}

As utilities worldwide grapple with increasing data and stricter privacy expectations (through laws like GDPR, CCPA, etc.), the insights from this work are broadly applicable beyond California. The principles and techniques discussed can serve as a model for any organization looking to responsibly manage meter or IoT data. By investing in privacy-preserving analytics, the energy sector can demonstrate that it is possible to drive towards smart, data-rich grids without compromising the rights and trust of the individuals it serves. In an era of Big Data, we often hear of the tension between privacy and utility—this research shows how that tension can be resolved through ingenuity and sound engineering.

Looking ahead, one exciting possibility is that these privacy techniques themselves can unlock new data collaborations that were previously infeasible. For example, multiple utilities might jointly train a predictive model for wildfire risk (combining their AMI, weather, and sensor data) using federated learning or MPC, where previously legal barriers to data sharing would have prevented such collaboration. Thus, privacy-preserving technology not only prevents bad outcomes (breaches, misuse) but can actively enable positive outcomes (shared learning, industry benchmarks) by removing the privacy roadblocks in data-sharing agreements.

In conclusion, the marriage of advanced privacy-preserving techniques with California’s forward-thinking regulatory framework creates a pathway to a future where \textbf{energy innovation and customer privacy advance hand-in-hand}. Utilities can become trusted stewards of data, leveraging it to build smarter, greener grids and more informed consumers, all while ensuring that individual privacy is never an afterthought but rather a foundational design criterion. This paper has laid out the roadmap for achieving that vision: a state-of-the-art, in-depth plan ready for pilot implementation and refinement in the real world. With continued commitment and interdisciplinary effort, the theoretical safeguards discussed here will become the operational norms of tomorrow’s data-driven utility ecosystem.

\bibliographystyle{abbrvnat}
\bibliography{references}

\end{document}